\documentclass[useAMS,usenatbib,12]{article}
\usepackage[dvips]{graphicx}
\title {\textbf{ How to Demonstrate the Voltage of a Charged Object in Physics Laboratory}}
\author {Raju Baddi,$^{1}$\\ $^1$National Center for Radio Astrophysics, TIFR, Ganeshkhind, P.O.Bag 3, PUNE 411007.\\}
\date{}
\begin{document}

\label{firstpage}

\maketitle

\begin{abstract}
Common Objects like a comb or a pen get charged when rubbed against something like human hair or garment clothing. 
Charged objects exhibit noticeable attractive or repulsive force lifting small pieces of paper or pushing/pulling a 
suspended light object charged with the same/opposite(uncharged) polarity respectively. This indicates the strong 
electrical nature of charged objects. Flashes due to spark between oppositely charged objects can be seen in total 
darkness. Implying a large potential difference between these charged objects which is not possible at lower voltages. 
This article describes a method to measure the voltage on commonly charged objects with respect to earth using simple 
instrumentation based on capacitors and CMOS voltmeter. Once the potential difference is known the average charge on 
the object can be calculated as well. The article also suggests a simple femto-farad capacitance meter for 
electrostatics work.     
\end{abstract}

\section{Introduction}
The electrical nature of charged objects is ubiquitously encountered by the rubbing of a comb against human hair. The 
comb is negatively charged while the human body positively. Such a comb can attract small pieces of paper lifting them 
from the ground over a short distance. Or when brought again near body hair sparks with crackling sound could be heard. 
This electrical behaviour can also be seen with other objects as well like an ordinary plastic pen rubbed against 
suitable garment. Flashes of spark can be seen in total darkness when suitable garments or clothing are rubbed against 
eachother. All this implies strong electrical nature or possibly large amount of charge on these objects which is 
normally not seen with electrical equipment which operate at a few volts to a few hundred volts. To assess the electrical 
condition of these charged objects(like a platic pen/narrow-cylinder or any other material brought to an appropriate form) 
a simple technique based on charging of capacitors is suggested which can be used for demonstration in a class room or 
to conveince one self the potential difference to which these objects are brought with respect to earth(considered to be 
at zero potential and which easily lends a small quantity of charge with a negligible change in its potential) in this 
charge exchange process. \\ 

When two capacitors with a large difference in magnitude are charged in series it is seen that, while one(smaller) 
preserves the almost total potential difference across the combination the other(larger) preserves the almost total charge 
on the combination at a negligible potential difference across it. Now if the capacitances of both the capacitors are known then by 
measuring the voltage on one of them the voltage on the other can be determined. As such its the voltage on the larger 
capacitor with a greater quantity of charge and smaller voltage that would be preferred. A measurement of its voltage(i.e 
across C$_{Large}$) multiplied by the ratio of the capacitors(C$_{Large}$/C$_{Small}$) would easily reveal the almost total 
potential difference across the combination. This very principle is applied here to determine the voltage on the charged 
object w.r.t earth. \\      
      
\section[]{The Apparatus}

\begin{figure}[here]
\begin{center}
\includegraphics[width=170mm,height=75mm,angle=0]{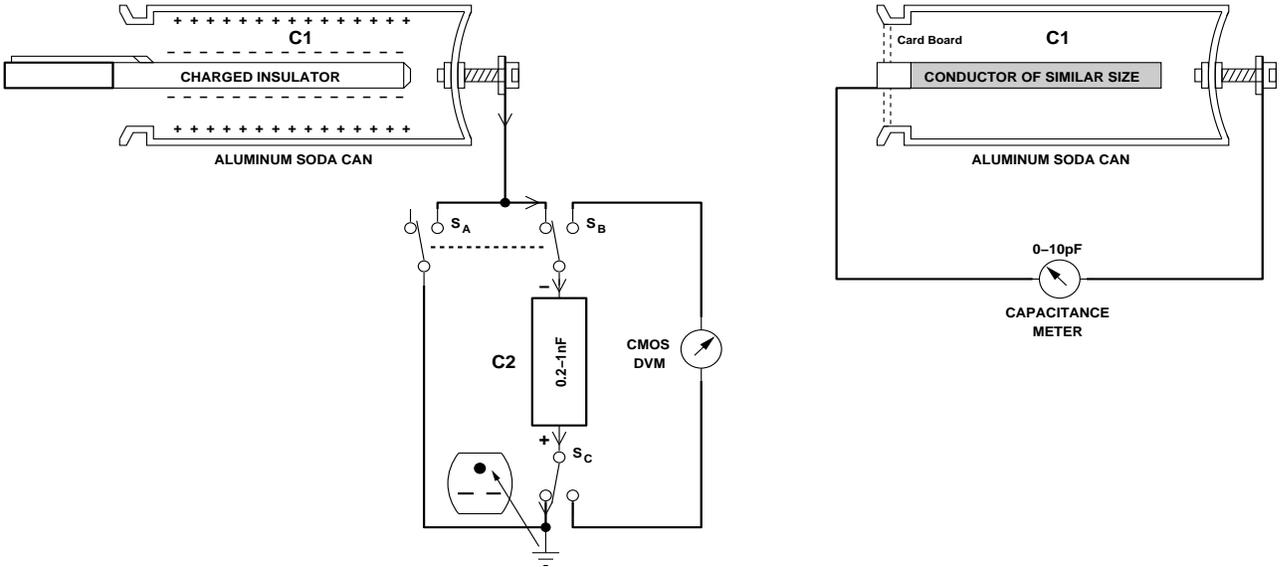}
\caption{ \textbf{Left:} The circuit for the experimental setup. Here a charged plastic pen is introduced into the cylinder which is a 
330ml aluminum soda can. The switches S$_A$, S$_B$ and S$_C$ are simultaneously operated and help conveniently connect/disconnect the 
reference capacitor in and out of the circuit. If you plan to use the high impedance CMOS voltmeter supplied with this article 
all one has to do is reset the voltmeter and throw the switch S to right and in a very short time($<$1 second) the voltage on the 
capacitor C2 can be determined indirectly. \textbf{Right:} Illustration of measurement of capacitance of C1 system with a picofarad meter. 
To immitate the form of the conductor one can roll Al-foil around the region of the object which is assumed to be charged. For 
distributed capacitance measurement make measurements by not connecting any of the terminals of C1 and then by connecting only 
one terminal at a time, note down these readings and subtract appropriately from the reading when both the terminals of the C1 
system are connected as shown in Figure 3. Further it should be cautioned to take care of any near by stray conductors near to the 
C1 system which can significantly alter the value. Perhaps a cross check measurement by swaping the terminals of C1 would be of 
help.}
\end{center}
\end{figure}

The apparatus is shown in figure 1, it consists of a cylinder(say a 330ml soda can) into which the object whose voltage 
w.r.t earth is to be determined is introduced. The object and the cylinder form a kind of capacitor system C1 which is 
connected in series with another reference capacitor C2 which is much larger(atleast 2 orders of magnitude) compared to C1. 
For the dimensions considered in this article C1 would be on the order of $\sim$ 1-5 pico farad(pF). So our reference capacitor 
can be about a few to several 100 pF upto 1nF(or perhaps even more) depending on the intensity of charge on the object. The 
other end of C2 is connected to earth which supplies any small quantity of charge and maintains its zero potential. When the 
charged object is inserted into the cylinder the potential difference between it and the cylinder surface will cause charges 
to flow from earth through the reference capacitor(in the process charging it) to the cylinder. The cylinder capacitor would 
receive as much charge as required to maintain its surface nearly at the same potential as earth as it is unhindered by the 
large capacitor C2. It should be mentioned here that as per our selection of its value the voltage drop developed across C2 
in the worst case is less than 10V and we guess the voltage on the charged 
object($\sim$10$^3$V) is orders of magnitude higher than this, essentially 10V(or lower) is negligible compared to it. So 
we have the metal cylinder at nearly the same potential as earth(0V) and the total charge that flowed from earth to C1 is 
stored in C2. In view of the CMOS voltmeter suggested in this article the value of C2 should be selected such that the voltage 
drop across it for a typical charge transfer is above 2.5V(if the -ve supply is 5V on the other hand if you choose 3V this would
be 1.2V) and less than 10V. \\
 
To measure the potential difference on the charged object insert it into the cylinder with switch in the left position and then 
throw it to right to measure the voltage on C2. The potential difference on the charged object w.r.t earth either positive 
or negative can be obtained as V$_{C2}\times$C2/C1. Care must be taken while connecting the capacitor C2 to the voltmeter so 
that it has the right polarity. It is possible to determine quickly the polarity of an intensely charged object using a 
single transistor. For this wire the transistor using a suitable power source with a bright LED in the collector and protective 
resistor(10k$\Omega$) in the base circuit, now bring the charged object near the base resistor and observe the LED as the object 
is moved back and forth. Else one can use the electronic electroscope described in [5]. \\
 
\section{Summary}
This article describes a simple experiment to demonstrate the voltage on a commonly charged object using simple low cost 
instrumentation. Author's measurement of capacitance(C1) were between 2-3pF and voltages upto $>$2500-1500V respectively with 
use of C2=200pF-1nF and suggested instrumentation. For example in one case C1$\sim$2.5pF, C2=1nF and V$_{C2}\sim$-4.5V which 
yields a voltage of (C2/C1)V$_{C2}\sim$-1800V for V$_{C1}$. It should be noted that the charged object looses its charge gradually
and hence intense charge measurements should be carried out as quickly as possible.

\section{Appendix}
This appendix supplies the required two instruments in the experiment. However its possible to use any other standard instrument 
available/suitable. If you are using a standard configuration for C1 such that the capacitance of the system can be calculated 
then one does not need a capacitance meter. However its availability makes measurement of capacitance of any configuration possible 
eliminating constraint on configuration of C1.
\subsection{High Impedance CMOS voltmeter}
\begin{figure}[here]
\begin{center}
\includegraphics[width=170mm,height=70mm,angle=0]{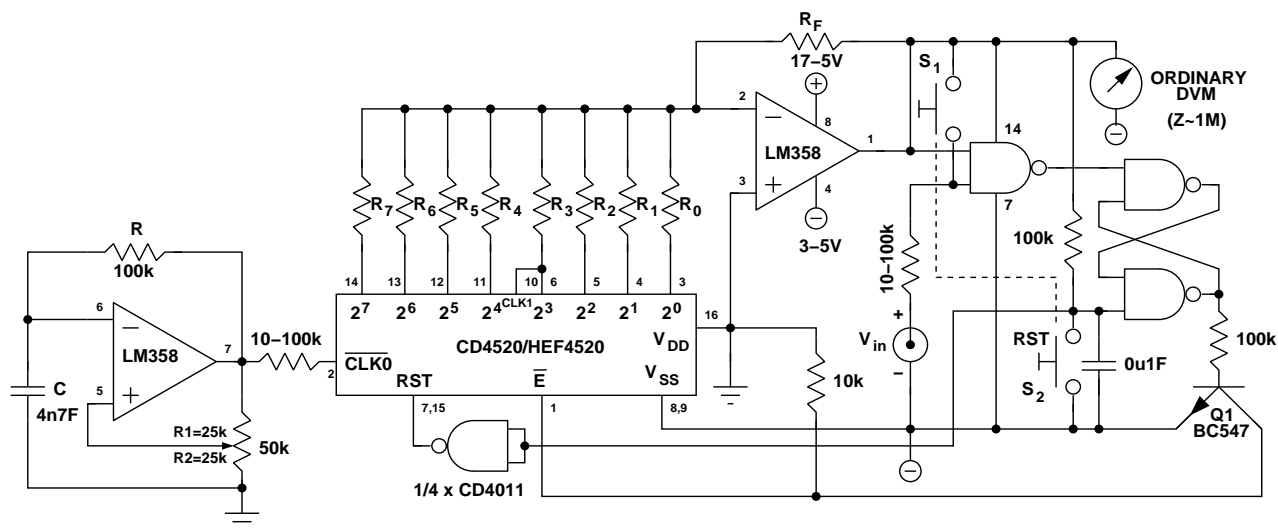}
\caption{Suggested CMOS voltmeter can estimate voltage across charged capacitors as small as 100pF without seriously modifying their 
charge content as the typical gate input current is $\sim$10$^{-11}$A[1]. However it should be cautioned that the supply voltage 
to the chip be always higher than the input voltage(V$_{in}$). Supply voltages are to be considered above and below ground respectively. 
R$_0$ through R$_7$ are to be selected such that they decrease by a factor of 2$^{n-1}$, suggested value for R$_0$ is 20-40k$\Omega$. 
It should be noted that pin 2 of the operational amplifier is more or less always at ground potential and so a constant voltage 
appears across each of the resistors and hence a constant current. Being turned on/off by the logic state of the counter(CD4520).
R$_F$ should be selected such that the integrated current(I$_{\Sigma}$=($2^n$-1$)\frac{|V_{ss}|}{R_0}$) flowing through it when all 
the resistors are on(i.e when the counter reads 00) is equal to the required maximum voltage(V$_{M}$=R$_F$I$_\Sigma$ ; 
R$_F$=V$_{M}$/I$_\Sigma$ ), in this case 15V above ground. It is also possible to use a CMOS OPAmp and build a conventional OPAmp 
based voltmeter.}
\end{center}
\end{figure}

This voltmeter(Figure 2) is based on estimating the applied voltage(V$_{in}$) by bringing it in proximity to the threshold voltage(V$_T$) 
of a common CMOS gate by varying the supply voltage to the package itself. 
In [2] this is done manually however here to determine the voltage on the capacitor in a very short time a digital to analog 
converter is used, together with a flip-flop and an oscillator. Once the supply voltage to the CD4011 reaches an appropriate value 
such that V$_{in}$ is just around V$_T$ the clocking of the counter stops and so does the voltage variation across CD4011. From the 
voltage across it($|$V$_{SS}|$+V$_{OPAmp}$) and by the knowledge of variation of V$_T$ with supply voltage one can determine V$_{in}$, 
which would be actually equal to V$_T$ corresponding to the supply voltage where the clocking stopped.  

\subsection{Femto Farad(fF) Capacitance Meter}
\begin{figure}[h]
\begin{center}
\includegraphics[width=140mm,height=90mm,angle=0]{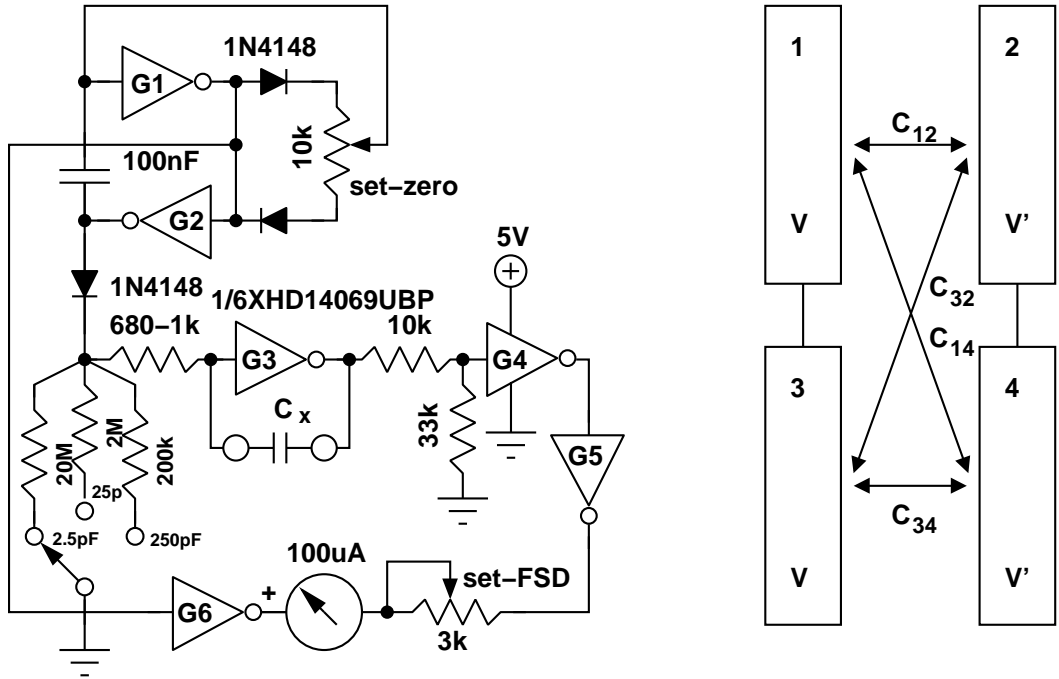}
\caption{Left: Suggested CMOS inverter gate based capacitance meter[3] which can measure capacitors on pico and femto scales. This 
is a very simple instrument which can be used in electrostatic experiments for the  measurement of femto and pico farad capacitance 
when appropriately calibrated. The instrument exhibits good linearity with capacitance over 1000's of division. One can also use a 
DVM(2000mV) in place of a 100$\mu$A meter. It should be mentioned here that the above instrument with DVM indicated nearly the same 
value as printed on the commercial capacitor from 500fF to 22pF. Right: Capacitance of distributed conductors, here conductors 1 \& 
3 and 2 \& 4 are at same potential V and V' respectively. When different conductors are connected together cross capacitance between 
different parts has to be taken into account during estimation of desired capacitance(here C$_{34}$).}
\end{center}
\end{figure}

The capacitance meter[3] suggested in Figure 3 can measure very low capacitances down to 10$^{-14}$F. The capacitance due to connected leads 
can also be nulled by using set-zero potentiometer. The range/calibration (F/div) of the instrument can be set with the help of the set-FSD 
resistor(moving coil meter) and also by varying the supply voltage(in this view the supply voltage has to be stable and strictly regulated).
The instrument is based on charging a capacitor with a constant low current between two different voltages to produce a delay in a closed 
loop duty cycle compensated 180$^o$ out of phase gate astable drive. It can measure capacitances on the order of 10$^{-14}$F(10 fF) to 
10$^{-9}$F(nF). It should be cautioned that measurement of distributed capacitance is not as straight forward as compact localized capacitance 
like that of a commercial capacitor or a twisted pair of wires. During measurement of distributed capacitance care must be taken to account 
for the cross capacitance terms between different parts of the conductors[4].

\label{lastpage}


\begin{thebibliography}{99}

\bibitem{} CD4011 data sheet from www.datasheetarchive.com .

\bibitem{} Logic gates form high-impedance voltmeter, EDN May 26 2011.

\bibitem{} A sub-pico farad capacitance meter, submitted to EDN magazine.

\bibitem{} Jackson J.D., Classical Electrodynamics, 3$^{rd}$ edition, 2004, page 43 Eqn (1.61).

\bibitem{} An Electronic Electroscope, submitted to EDN magazine.

\end{thebibliography}
\end{document}